# Recent Spin Rate Measurements of 4U 1907+09

Ş. Şahiner*, S. Ç. İnam† and A. Baykal*

*Physics Department, Middle East Technical University, 06531 Ankara, Turkey
†Electrical and Electronics Engineering Department, Başkent University, 06530 Ankara, Turkey

**Abstract.** In this study, X-ray spectral and pulse timing analysis of the high mass X-ray binary (HMXB) pulsar 4U 1907+09, based on the observations with *RXTE* are presented. Spin rate measurements indicate a new spin-down episode with a rate close to the previous steady spin-down rate. Orbital phase resolved spectroscopy reveals that the Hydrogen column density varies through the orbit reaching to its maximum value just after periastron. A slight spectral softening with increasing luminosity is aslo observed.

## INTRODUCTION

The HMXB system 4U 1907+09 consists of an X-ray pulsar accreting material from its blue supergiant companion. The orbital period of the system is ∼8.38 d [1], with two phase-locked maxima; the primary corresponding to periastron and the secondary to apastron [2]. The spin period of the pulsar was first measured as ∼437 s [3]. The historical measurements of pulse period confirmed that the source exhibited a steady spin down rate until 1998 [2]. Afterwards, in 2001 a significant decrease of spin down rate was reported by Baykal et al. [4] and in 2004 a complete trend reversal to spin up was reported by Fritz et al. [5]. Recent measurements performed by İnam et al. [6] and continuing with this work reveals a new spin down episode.

*RXTE* monitoring observations of 4U 1907+09 with the proposal IDs 93036 and 94036 are held on between June 2007 and December 2009. There are 63 pointed observations each having an average exposure time ∼2 ks. The data products of *PCA* detectors are reduced for the analysis.

## RESULTS & DISCUSSION

Background subtracted non-dipping light curves, corrected to the barycenter of the Solar system and to the binary orbital motion, are analyzed to perform the pulse timing measurements. Each ∼2 ks long light curve segment is folded on a statistically independent trial period [7]. A template pulse profile is constructed from the first three observations with a total exposure of ∼6 ks by folding the data on the period giving the maximum $\chi^2$. The template pulse and independent sample pulses are represented as Fourier series and compared with each other harmonic by harmonic [8], consequently the sample pulse phase is determined by the pulse phase offset $\delta\phi$ between the sample and the master pulse. In order to estimate pulse frequency derivatives, the pulse phase offsets

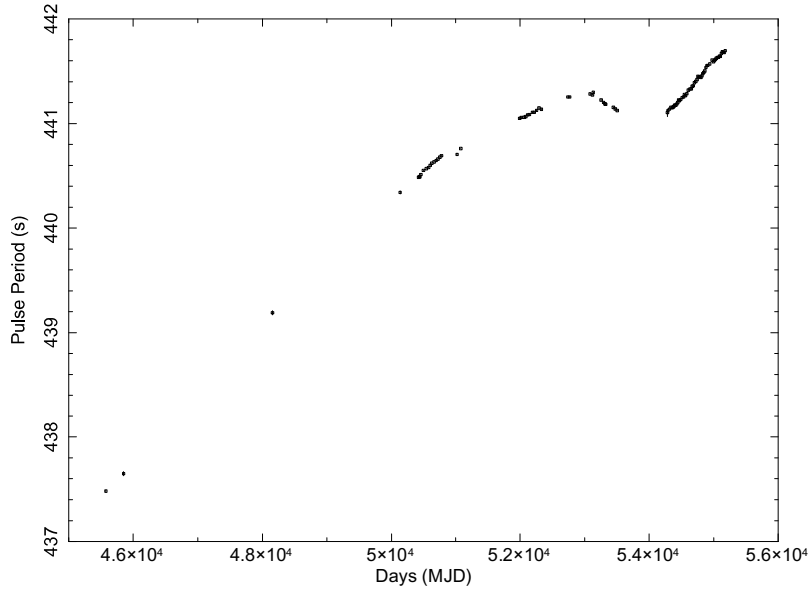

**FIGURE 1.** Pulse period history of 4U 1907+09. Recent spin-down trend of the source is evident from MJD 54281 onwards.

calculated from the cross-correlation analysis are fitted to the quadratic polynomial in Eq. 1;

$$\delta\phi = \phi_o + \delta v(t-t_o) + \frac{1}{2}\dot{v}(t-t_o)^2 \qquad (1)$$

where, $t_o$ is the mid-time of the observation, $\phi_o$ is the phase offset at $t_o$, $\delta v$ is the deviation from the mean pulse frequency, $\dot{v}$ is the pulse frequency derivative. In order to demonstrate the pulse period evolution of the source, 53 pulse periods are estimated by calculating individual derivatives form each pair of pulse phases. Combining these results with the previous measurements, pulse period history of 4U 1907+09 is plotted in Fig. 1.

3-25 keV spectrum of each non-dip observation is individually modeled to investigate the variation in the spectral parameters through the binary orbit. The basic model comprises a simple power law with high energy cutoff and photoelectric absorption. Deviation of residuals around 20 keV confirmed the presence of cyclotron absorption feature at ∼19 keV previously reported by Cusumano et al. [9]. Orbital dependence of Hydrogen column density ($n_H$) is evident from the results. The base value of $n_H$ increases up to a maximal value just after the periastron passage and it remains at high values until the apastron, where it reduces to its base value again. The orbital dependence of $n_H$ approves that the location of the absorbing material is the accretion flow. Preservation of the maximal absorption from periastron to apastron implies that the gas stream from the companion trails behind the pulsar until the apastron. There is no significant orbital dependence of the power law photon index. Furthermore, the photon index slightly increases with increasing flux, showing that the spectra of 4U 1907+09 softens during high flux episodes. This relation may be a probe to formation of an accretion disk since soft photons mostly arise from the reproduction of photons at the disk. The results show no correlation between column density and flux, which is contradicting. In fact, a reduction of opacity is expected due to an increase in ionization with increasing flux. This inconsistency may be arising from the variability of mass accretion rate.

Torque reversal is a rare event for 4U 1907+09, in contrast to most wind-fed pulsars which commonly undergo alternating episodes of spin-up and spin-down (e.g. Vela X-1, Cen X-3). Decade-long distinct episodes are also observed in systems such as 4U 1626-67 and GX 1+4 [10]. According to Ghosh and Lamb [11] model of accretion torques, a net negative torque resulting in a spin-down may arise either from an increase in magnetic field or decrease in mass accretion rate. However such changes in the system should be accompanied with changes in the X-ray flux and spectral characteristics. 4U 1907+09 is rather a challenging example; because no significant flux or spectral variations related to the changes in the spin rate, have been observed.

Recent models considering retrograde disks, warping and precessing disks, angular velocity transitions in the disk etc. also require changes in the X-ray flux and spectral characteristics, therefore they cannot be applied for 4U 1907+09. Only the model of Perna et al. [12], considering a tilted magnetic axis, can explain torque reversals without a need of change in the mass accretion rate. This model shows that; different regions can occur in the disk due to an asymmetric magnetospheric boundary, such that the propeller effect is locally at work in some regions, while accretion from other regions is possible. Cyclic torque reversal episodes may be generated by this accretion geometry proposed for persistent prograde accretion discs; however the nature of the accretion disc in 4U 1907+09 is of transient nature.

## ACKNOWLEDGMENTS

We acknowledge the support from TÜBİTAK, the Scientific and Technological Research Council of Turkey through the research project TBAG 109T748 and EU FP6 Transfer of Knowledge Project "Astrophysics of Neutron Stars" (MTKD-CT-2006-042722).